\RequirePackage{lineno}

\documentclass[3p,english,times,procedia]{elsarticle}
\usepackage{nupha_ecrc}
\usepackage{color}
\usepackage{multirow}
\usepackage{amsmath}
\usepackage{graphicx}
\usepackage{babel}
\usepackage[normalem]{ulem}
\newcommand {\snn}	{\sqrt{s_{_{\rm NN}}}}

\newcommand {\dvvach}	{\Delta v_{2}(A_{\rm ch})}
\newcommand {\dvvvach}	{\Delta v_{3}(A_{\rm ch})}
\newcommand {\dvnach}	{\Delta v_{n}(A_{\rm ch})}

\newcommand {\ach} {A_{\rm ch}}

\volume{00}

\firstpage{1}

\journalname{Nuclear Physics A}

\runauth{}

\jid{nupha}

\jnltitlelogo{Nuclear Physics A}

\usepackage{amssymb}

\usepackage[figuresright]{rotating}

\begin{document}

\begin{frontmatter}



	\dochead{XXVIIIth International Conference on Ultrarelativistic Nucleus-Nucleus Collisions\\ (Quark Matter 2019)}

	\title{Importance of non-flow background on the chiral magnetic wave search}

	\author[label1,label2]{Hao-jie Xu}
	\author[label2]{Jie Zhao}
	\author[label2]{Yicheng Feng}
	\author[label1,label2]{Fuqiang Wang}
	\address[label1]{School of Science, Huzhou University, Huzhou, Zhejiang 313000, China}
	\address[label2]{Department of Physics and Astronomy, Purdue University, West Lafayette, Indiana 47907, USA}

	\begin{abstract}
		An observable sensitive to the chiral magnetic wave (CMW) is the charge asymmetry dependence of the $\pi^{-}$ and $\pi^{+}$ anisotropic flow difference, $\Delta v_{n}(A_{\rm ch})$. 
		We show that, due to non-flow correlations, the flow measurements by the Q-cumulant method using all charged particles as 
		reference introduce a trivial linear term to $\Delta v_{n}(A_{\rm ch})$.
		The trivial slope contribution to the triangle flow difference $\Delta v_{3}(A_{\rm ch})$ can be negative 
		if the non-flow is dominated by back-to-back pairs. This can explain the observed negative  $\Delta v_{3}(A_{\rm ch})$ slope in the preliminary STAR data.
		We further find that the non-flow correlations give rise to additional backgrounds to the slope of $\Delta v_{2}(A_{\rm ch})$ 
		from the competition among different pion sources and from the larger multiplicity dilution to $\pi^{+}$ ($\pi^{-}$) at positive (negative) $A_{\rm ch}$. 
	\end{abstract}


\end{frontmatter}


\section{Introduction}
\label{sec:introduction}

The interplay between the chiral magnetic effect and the chiral separation effect can 
lead to a gapless collective excitation, a phenomenon called the chiral magnetic wave (CMW)~\cite{Kharzeev:2010gd,Burnier:2012ae}. 
The CMW could introduce an electric quadrupole moment, 
giving opposite contributions to the $\pi^{+}$ and $\pi^{-}$ elliptic flow anisotropies ($v_{2}$) dependent of the charge asymmetry ($\ach=\frac{N_{+}-N_{-}}{N_{+}+N_{-}}$)~\cite{Burnier:2012ae} 
\begin{equation}
	v_{2}\{\pi^{\pm}\} =  v_{2}^{\rm base} \mp \frac{r(\pi^{\pm})}{2}\ach.
\end{equation}
The CMW-sensitive slope parameters ($r$) measured by the STAR, ALICE and CMS collaborations qualitatively agree with the expectation from the CMW~\cite{Adamczyk:2015eqo,Adam:2015vje,Sirunyan:2017tax}.
The data can also be qualitatively explained by non-CMW mechanisms, such as 
the Local Charge Conservation (LCC)~\cite{Bzdak:2013yla} and the effect of isospin chemical potential~\cite{Hatta:2015hca}. 
We will show in these proceedings that non-flow correlations can also cause $\ach$-dependent $\pi$ flows.
We demonstrate~\cite{Xu:2019pgj} that
the non-flow correlations can give both trivial and non-trivial contributions to the slope parameters of $\dvnach\equiv v_{n}^{\pi^-}(\ach)-v_{n}^{\pi^+}(\ach)$, 
where $n=2$ (elliptic flow) and $n=3$ (triangle flow).

\section{Trivial non-flow contributions to $v_{n}(\ach)$}
\label{sec:trivial}

Using the  $Q_{n}$-vector $Q_{n}=\sum_{i=1}^{M}e^{in\varphi_{i}}$, 
the anisotropic flow of particles of interest (POI, $\pi^{\pm}$ in this study) can be calculated by
$v_{n}^{\pi^{\pm}}\{2\}=\frac{d_{n}\{2;\pi^{\pm}h\}}{\sqrt{c_{n}\{2\}}}$ 
with $d_{n}\{2\}\equiv \langle\langle2'\rangle\rangle = \frac{\sum_{i} w_i\langle2'\rangle_{i}}{\sum_i w_i}$ 
, $\langle2'\rangle_i \equiv \frac{q_{n,i}Q_{n,i}^{*}}{m_iM_i}$, and $\sqrt{c_{n}\{2\}}$ is the flow of reference particles (REF).
Here $w_i=m_{i}M_{i}$, $(m_i,q_{n,i})$ and $(M_i,Q_{n,i})$ are the (multiplicity, Q-vector)
of POI and REF, respectively.

With all charged hadrons as REF, as typically done in data analysis, the two-particle cumulant can be rewritten into~\cite{Xu:2019pgj}
\begin{equation}
	d_{n}\{2;\pi^{\pm}h\} =\frac{d_{n}\{2;\pi^{\pm}h^{+}\}+d_{n}\{2;\pi^{\pm}h^{-}\}}{2}  
		+\frac{d_{n}\{2;\pi^{\pm}h^{+}\}-d_{n}\{2;\pi^{\pm}h^{-}\}}{2}\ach. \label{eq:trivial}
\end{equation}
The second term on r.h.s of Eq.~(\ref{eq:trivial}) is proportional to $\ach$ and opposite in sign for $\pi^{+}$ and $\pi^{-}$.
This will directly give a trivial contribution to the CMW-sensitive slope parameter. 
It vanishes if the correlations are due to flow only because in this case $d_{n}\{2;\pi^{\pm}h^{+}\}=d_{n}\{2;\pi^{\pm}h^{-}\}$.
However, non-flow is present in experimental data and 
differs between like-sign and unlike-sign pairs, so the trivial term is finite. 

\begin{figure*}[tbp]

	\begin{centering}
		\includegraphics[scale=0.3]{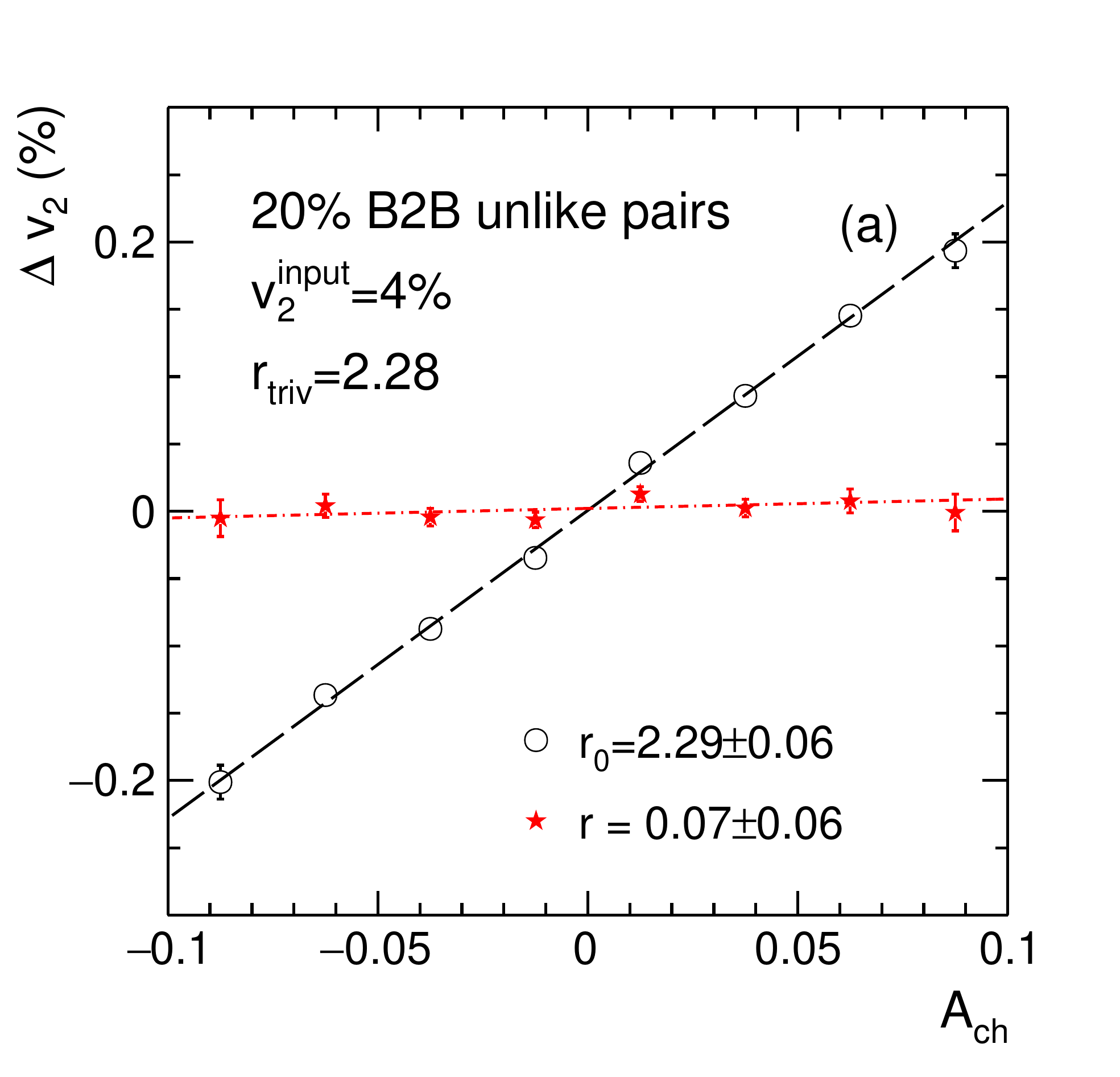}\includegraphics[scale=0.3]{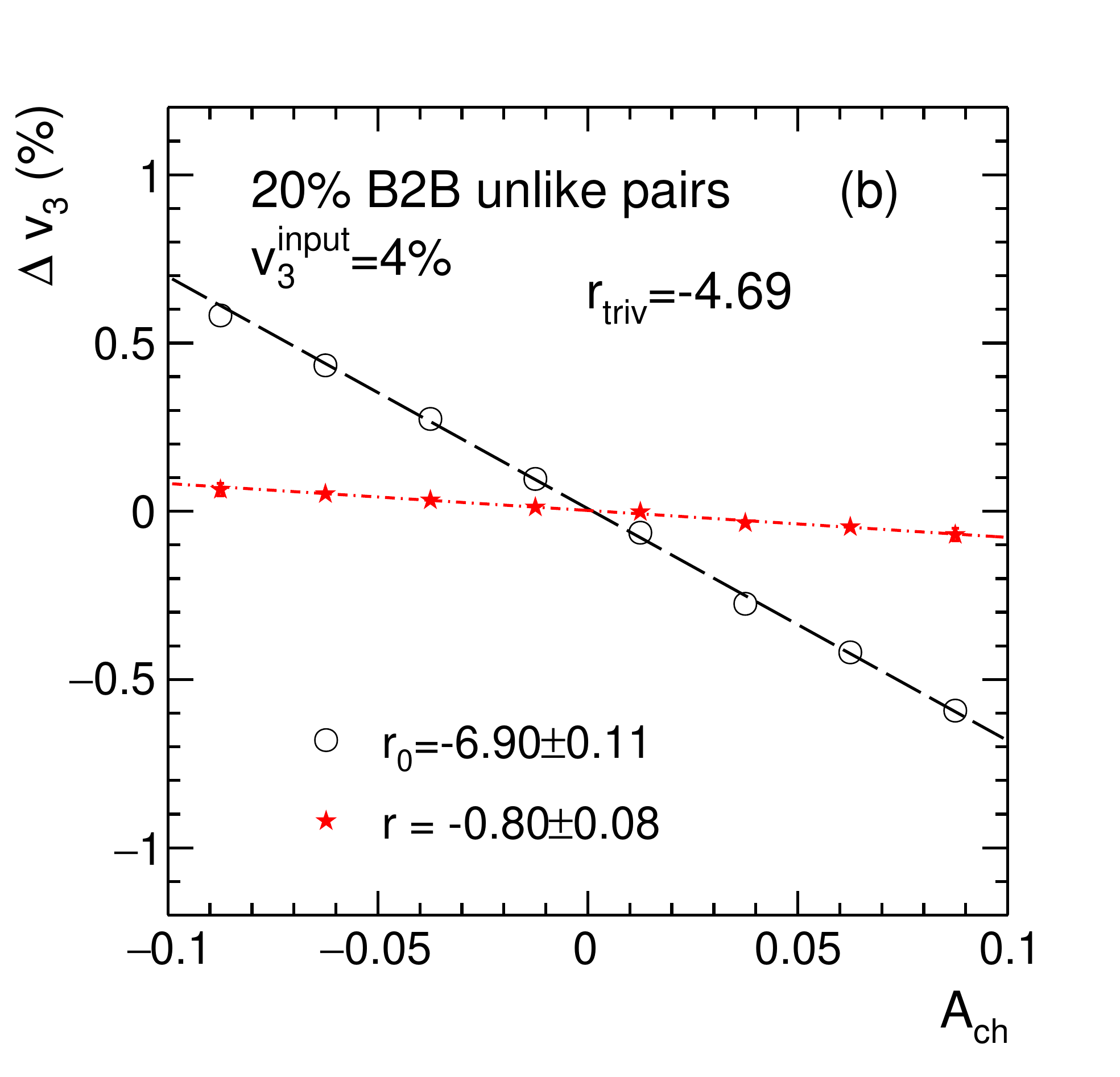} 
	\par\end{centering}
	\caption{(Color online)
	A Monte Carlo model demonstration of the trivial term, arising from  back-to-back (B2B) unlike-sign pair non-flow correlations, due to the net effect of non-flow 
	difference between like-sign and unlike-sign pairs and using all charged particles as REF: (a) $\dvvach$ and (b) $\dvvvach$.
	Results before and after eliminating the trivial term are shown by open circles and filled stars, respectively.
	\label{fig:toy2}}
\end{figure*}

The STAR preliminary results indicate a negative slope for $\dvvvach$ in central and peripheral collisions~\cite{Shou:2018zvw}.
A negative trivial slope can easily arise from back-to-back pairs of particles. We illustrate this using a Monte Carlo model.
We generate $\pi^{+}$ and $\pi^{-}$ with Poisson multiplicity fluctuations in each event. 
The $p_{T}$ spectra correspond to the measured data in the $30\mbox{-}40\%$ centrality Au+Au collisions at $\snn=200$ GeV~\cite{Adamczyk:2015lme,Zhao:2017nfq};
The $\eta$ spectra are parameterized as in Ref.~\cite{Alver:2010ck}.
The mean multiplicity of charged hadrons is set to $380$ in $|\eta|<1$ with $p_{T}>0.15$ GeV/c.
To introduce a non-flow correlation difference between like-sign and unlike-sign pairs, we
force, on average, $20\%$ of the multiplicity in a given event to come from $\pi^{+}\pi^{-}$ pairs with back-to-back azimuthal angles for the two pions.
A constant elliptic flow $v_{2}=4\%$ (triangle flow $v_{3}=4\%$) is used to generate the azimuth
angle of those pairs as well as the rest $80\%$ $\pi^{+}$ and $\pi^{-}$. 
The results are shown in Fig.~\ref{fig:toy2}.
The slope of the trivial term, dubbed the trivial slope $r_{\rm triv}$, is calculated by
$r_{\rm triv}(\pi^{\pm}) =  \frac{d_{n}\{2;\pi^{\pm}h^{+}\}-d_{n}\{2;\pi^{\pm}h^{-}\}}{2\sqrt{c_{n}\{2\}}}$ (c.f. Eq.~(\ref{eq:trivial})).
The slope parameter without removing the trivial term is denoted as $r_{0}$.
The back-to-back pairs contribute a positive trivial slope to $\dvvach$ shown in Fig.~\ref{fig:toy2}(a) and a negative trivial slope to $\dvvvach$ shown in Fig.~\ref{fig:toy2}(b).

Non-flow differences are present between like-sign and unlike-sign pairs in real collisions, and not much can be done to eliminate these non-flow differences.
In order to eliminate the trivial linear $\ach$ term, one can use hadrons of a single charge sign instead of all charged hadrons as REF.
One may use positive and negative particles as REF separately to obtain $v_{n}^{\pi}\{2;h^{+}\}$ and $v_{n}^{\pi}\{2;h^{-}\}$, 
and then take an average
\begin{equation}
	\bar{v}_{n}^{\pi} \equiv \frac{v_{n}^{\pi}\{2;h^{+}\} + v_{n}^{\pi}\{2;h^{-}\}}{2}.
	\label{eq:avgvn}
\end{equation}
The $\dvnach$ dependences obtained using this technique are shown in Fig.~\ref{fig:toy2} by the red stars.
Indeed, the slope is zero for $v_2$ as expected, because there is no other physics in our toy model that would introduce a non-zero slope.
It is interesting to note, however, that the $r$ slope for $\dvvvach$ does not completely vanish for back-to-back non-flow pairs, as shown in Fig.~\ref{fig:toy2}(b) by the red stars.
The reason is due to a competition between two sources of pions, the paired pions and unpaired pions, 
to be discussed in the next section.

%

\begin{figure*}[tbp]

	\begin{centering}
		\includegraphics[scale=0.4]{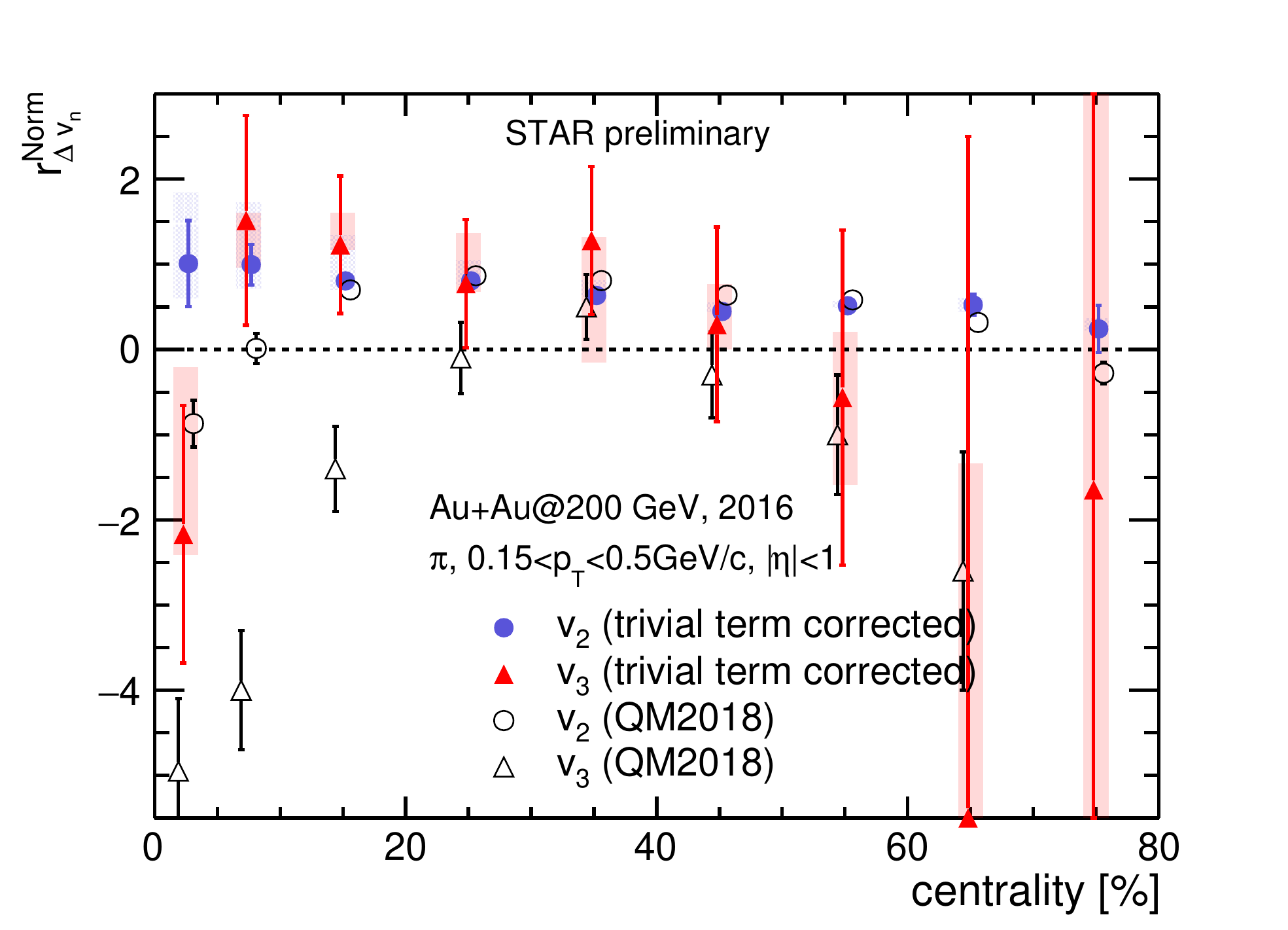} 
	\par\end{centering}
	\caption{(Color online)
	The STAR preliminary results on the slope parameters of the normalized $\dvvach$ and $\dvvvach$, i.e., $2\dvnach/(v_{n}(\pi^{+})+v_{n}(\pi^{-}))$
	before and after eliminating the trivial term. The figure is taken from Ref.~\cite{Xu:2019qm}.
	\label{fig:data}}
\end{figure*}

Preliminary STAR results showed significant negative slopes for $\dvvvach$~\cite{Shou:2018zvw}.
The negative slopes were taken as a strong evidence in favor of the CMW inferred from the $\dvvach$ data.
After eliminating the trivial slope following our methodology,
the normalized $\dvvvach$ slopes are now consistent with positive values (1.5~$\sigma$ above zero for $20-60\%$ centrality), 
and similar to the normalized $\dvvach$ slopes in terms of the relative magnitudes~\cite{Xu:2019qm}.
The new STAR preliminary results are shown in Fig.~\ref{fig:data}~\cite{Xu:2019qm}.

\section{Non-trivial non-flow contributions to $v_{n}(\ach)$}
\subsection{Competition between different pion sources}
The non-flow correlations can also give non-trivial contributions to $\dvnach$.
This is indeed shown in the $v_{3}$ results in the previous section. The underlying reason is the competition between different pion sources.
We now demonstrate this by using a two-component model,
i.e., primordial pions (denoted by subscript `P') and pions from resonance decays (denoted by `D').
We have 
$v_{n\pm} =\frac{N_{P\pm}v_{n,P\pm}+N_{D\pm}v_{n,P\pm}}{N_{D\pm}+N_{P\pm}}$,
	$\ach  = (1-\epsilon)A_{P} + \epsilon A_{D}$, 
$A_{P}=(N_{P+}-N_{P-})/(N_{P+}+N_{P-})$, 
$A_{D}=(N_{D+}-N_{D-})/(N_{D+}+N_{D-})$, 
	$\epsilon \equiv\frac{N_{D+}+N_{D-}}{N_{P+}+N_{P-}+N_{D+}+N_{D-}}$, 
where $N_{P\pm}$ ($N_{D\pm}$) and $v_{n,P\pm}$ ($v_{n,D\pm}$) are the
multiplicity and anisotropic flow of primordial (decay) $\pi^{\pm}$.

Assuming, without loss of generality, $v_{n,P}=v_{n,P+}=v_{n,P-}$ and $v_{n,D}=v_{n,D+}=v_{n,D-}$, 
independent of charge asymmetry, we have~\cite{Xu:2019pgj}
\begin{equation}
	\Delta v_{n}\simeq 2\epsilon(1-\epsilon)(A_{D}-A_{P})(v_{n,P}-v_{n,D})=\frac{2\epsilon(\epsilon\sigma_{D}^{2}-(1-\epsilon)\sigma_{P}^{2})(v_{n,P}-v_{n,D})}{(1-\epsilon)(\epsilon^{2}\sigma_{D}^{2}+(1-\epsilon)^{2}\sigma_{P}^{2})}\ach\equiv r^{2C}\ach. \label{eq:twocomponent}
\end{equation}
Here we have assumed the event-by-event distributions of $A_P$ and $A_D$ are both normal distributions, 
i.e., $\mathcal{N}(\mu_{P},\sigma_{P}^{2})$  and $\mathcal{N}(\mu_{D},\sigma_{D}^{2})$ in a charge-neutral system.

The slope $r^{2C}$ from the two-component (2C) model is clearly non-zero if $\sigma_{P}^{2}\neq \epsilon\sigma_{D}^{2}/(1-\epsilon)$ and $v_{n,P}\neq v_{n,D}$.
The root reason is that the relative fractions of pions from different sources depend on the event-by-event $\ach$ value (because they contribute to $\ach$ differently),
therefore the average $v_{2}$ from multiple sources, which have different $v_{2}$'s, will depend on $\ach$.

We have used two ``flow'' sources in the above derivation. However, this also applies to the competition between flow and non-flow
contributions to the observed $\dvnach$. 
This is the reason for the non-zero slope in Fig.~\ref{fig:toy2}(b) even after eliminating the trivial term,
because the ``$v_{3}$'' from the back-to-back pairs is by definition zero, which differs from 
the single pion $v_{3}$, 
even though the back-to-back pairs are generated with the same $v_{3}$ modulation.
Such a problem is not present for $v_{2}$. 
We have tested $v_{2}$ using two different input $v_{2}$'s  for single and paired pions, 
and also found a non-zero slope parameter.

\subsection{Like-sign non-flow correlations}
The non-flow correlations from like-sign pairs can also
introduce a non-zero slope parameter.
We modify our non-flow Monte Carlo model to generate like-sign pairs by forcing $20\%$ 
of $\pi^{+}$ (and $\pi^{-}$) to be paired as $\pi^{+}\pi^{+}$  (and $\pi^{-}\pi^{-}$) with the same azimuth.
All other parameters of the model are unchanged.
The resulting  $\dvvach$ has a positive slope $r=1.63\%$. 
This is due to the dilution effect: when more $\pi^{+}$ are counted resulting in a positive $\ach$, the $\pi^{+}\pi^{+}$ non-flow correlation is more diluted while 
the $\pi^{-}\pi^{-}$ non-flow is less diluted, resulting in a large $v_{2}$ for $\pi^{-}$ than for $\pi^{+}$. 
This is different in the unlike-sign case, where the dilution effect is identical for $\pi^{+}$ and $\pi^{-}$.

\section{Summary}
The charge asymmetry ($\ach$) dependent pion elliptic flow difference $\dvvach$ is a sensitive observable to the chiral magnetic wave (CMW).
In these proceedings, we first demonstrate that the flow measurements can automatically introduce a trivial linear-$\ach$ dependence if 
(1) there exists non-flow difference between like-sign and unlike-sign pairs and 
(2) hadrons of both charge sings are used as reference particles in the two-particle cumulant flow measurements.
Using a Monte Carlo model, we find that back-to-back unlike-sign pair non-flow correlations contribute a positive trivial slope to $\dvvach$ and a negative trivial slope to $\dvvvach$.
New data analysis indicates that the trivial contribution is the dominate reason for the large negative slope of $\dvvvach$
in the previous STAR preliminary results (see Fig. \ref{fig:data}).

We further find that the competition among multiple $\pi$ sources can introduce a non-trivial linear-$\ach$ term.
This effect is sensitive to the differences in multiplicity fluctuations and anisotropic flows of those sources, and arises from the $\ach$-dependent relative contributions of pions from those sources. 
We also find that the  non-flow between like-sign pairs gives a positive slope to $\Delta v_{2}(A_{\rm ch})$ 
because of the larger multiplicity dilution effect to $\pi^{+}$ ($\pi^{-}$) at positive (negative) $A_{\rm ch}$.




\

{\em Acknowledgments:} This work is supported in part by the National Natural Science Foundation of 
China (Grant Nos. 11905059, 11847315, 11947410) and the U.S. Department of Energy (Grant No. DE-SC0012910).  
HX acknowledges financial support from the China Scholarship Council.

\end{document}